\documentclass[11pt,a4paper]{article}

\usepackage[includeheadfoot,
            marginratio={1:1,2:3}, 
            width=412pt, 
            height=688pt,]{geometry}

\usepackage{amsmath,mathrsfs}
\usepackage{amsfonts}
\usepackage{amssymb}
\usepackage{stmaryrd}
\usepackage{graphicx}
\usepackage{caption}
\usepackage{cite}
\usepackage{etoolbox}
\usepackage{makeidx}
\usepackage{authblk} 
\usepackage{hyperref}
\hypersetup{
pdftitle={},%
pdfauthor={},%
pdfsubject={},%
pdfkeywords={},%
colorlinks=true,%
linkcolor=blue,%
citecolor=red,%
linktocpage=true,%
pageanchor=true
}
\DeclareMathAlphabet{\mathpzc}{OT1}{pzc}{m}{it}

\usepackage{empheq}
\usepackage{paralist}

\newcommand{\nc}{\newcommand}
\nc{\lb}{\llbracket}
\nc{\rb}{\rrbracket}
\nc{\gl}{\llbracket}
\nc{\gr}{\rrbracket}

\newcommand{\eq}[1]{\begin{equation}
                     \begin{split} #1 \end{split}
                     \end{equation}}

\newcommand{\be}{\begin{equation}}
\newcommand{\ee}{\end{equation}}
\newcommand{\R}{\mathbb R}

\newcommand{\C}{\mathbb C}
\def\beqa{\begin{eqnarray}}
\def\eeqa{\end{eqnarray}}
\newcommand{\eqn}[1]{(\ref{#1})}

\newcommand\DER{{\text{\textup{Der}}}}
\newcommand\modul{{\mathbb M}}
\newcommand\caZ{{\mathcal Z}}
\def\beqa{\begin{eqnarray}}
\def\eeqa{\end{eqnarray}}
\def\bean{\begin{eqnarray*}}
\def\eean{\end{eqnarray*}}
\newcommand{\del}{\partial}

\allowdisplaybreaks[2]
\numberwithin{equation}{section}
\title{\textbf{A novel approach to non-commutative gauge theory}\vspace{0.5cm}}
\date{}

\author[1,2]{Vladislav G. Kupriyanov }
\author[3,4]{Patrizia Vitale}
\affil[ ]{}
\affil[1]{\textit{\footnotesize CMCC-Universidade Federal do ABC, Santo Andr\'e, SP, 
Brazil}}
\affil[2]{\textit{\footnotesize Phisics Department, Tomsk State University, Tomsk, Russia}}
\affil[3]{\textit{\footnotesize Dipartimento di Fisica ``E. Pancini'', Universit\`a di Napoli Federico II, Complesso Universitario di Monte S. Angelo Edificio 6, via Cintia, 80126 Napoli, Italy.}}
\affil[4]{\textit{\footnotesize INFN-Sezione di Napoli, Complesso Universitario di Monte S. Angelo Edificio 6, via Cintia, 80126 Napoli, Italy.}}
\affil[ ]{}
\affil[ ]{\footnotesize e-mail: \texttt{vladislav.kupriyanov@gmail.com, patrizia.vitale@na.infn.it}}

\begin{document}
\maketitle

%%%%%%%%%%

\begin{abstract}
\baselineskip=12pt
\noindent 

We propose a field theoretical model defined on  non-commutative space-time with non-constant non-commutativity parameter $\Theta(x)$, which satisfies two main requirements: it is gauge invariant and reproduces in the commutative limit, $\Theta\to 0$, the standard $U(1)$ gauge theory. We work in the slowly varying field approximation where higher derivatives terms in the star commutator are neglected and the latter  is approximated by the Poisson bracket, $-i[f,g]_\star\approx\{f,g\}$. We derive an explicit expression for both the NC deformation of Abelian gauge transformations which close the algebra $[\delta_f,\delta_g]A=\delta_{\{f,g\}}A$, and the NC field strength ${\cal F}$,  covariant under these transformations, $\delta_f {\cal F}=\{{\cal F},f\}$. NC Chern-Simons equations 
are equivalent to the requirement that the NC field strength,  ${\cal F}$, should vanish identically. Such equations are non-Lagrangian. The NC deformation of Yang-Mills theory is obtained from the gauge invariant action, $S=\int {\cal F}^2$. As guiding example, the case of  $su(2)$-like  non-commutativity,  corresponding to  rotationally invariant NC space, is worked out in detail. 
\end{abstract}

%\end{titlepage}

%\setcounter{page}{2}

\newpage

%{\baselineskip=12pt
\tableofcontents
%}

\bigskip

\renewcommand{\thefootnote}{\arabic{footnote}}
\setcounter{footnote}{0}

\section{Introduction}
Noncommutative gauge theories have been widely studied in the past years mixing successes and defeats. It is of common knowledge that, while space-time noncommutativity represents a natural resolution of the clash between general relativity and quantum mechanics in strong gravitational fields (see for example \cite{Doplicher}), it doesn't give rise to  well defined quantum field theories, which are generically affected by the so called  UV/IR  mixing \cite{mixing}, except for a few models with very special features.  It is out of the scope of this paper to present a comprehensive review of the subject. We shall just focus on the aspects which shall be addressed here. 

 For the purposes of the paper, the``classical picture" which we refer to,    is that of a noncommutative  theory of gauge and (when included) matter fields, 
 % \textcolor{blue}{[here we should be more careful since later we say that in this paper we consider the pure gauge theory without matter fields]} 
% interacting with gauge fields, 
which is   described in terms a noncommutative algebra $(\mathcal{A}, \star)$ representing space-time, a right $\mathcal{A}$-module,  $\mathbb{M}$, representing matter fields, a group of unitary automorphisms of $\mathbb{M}$ acting on fields from the left, representing $U(N)$ gauge transformations.\footnote{Notice however that in   the paper we shall  only consider pure gauge theories.}
%, $\mathcal{G}= \{g: \mathcal{A}\rightarrow \mathfrak{g}\}$. 
In such a framework, the  dynamics of fields is  described by means of  a natural differential calculus based on  derivations  of the NC algebra \cite{diffcalc,dbv}. Moreover, the  gauge connection is  the standard noncommutative analog of the Koszul notion of connection \cite{dbv, wallet}. (See Appendix \ref{appa} for a brief review of the latter approach. For a physically inspired perspective see the pioneering work   \cite{Wess}.)

Therefore,  it is evident that, in the classical  framework, one major problem is to have  a well defined differential calculus, namely, an algebra of $\star$-derivations  of $\mathcal{A}$ such that
\be
D_a (f\star g)= D_a f\star g+ f\star D_a g.
\ee
For constant noncommutativity, assuming $\Theta$ to be non-degenerate, the latter are successfully realised by star commutators 
\be 
D_a f= (\Theta^{-1})_{ab}[x^b, f]_\star\stackrel{\Theta\rightarrow 0}{\longrightarrow} \del_a f \label{der}
\ee
thus reproducing the correct commutative limit. 
For  coordinate dependent $\Theta(x)$ the situation is much more complicated. Lie algebra type star products, 
\be
[x^j, x^k]_\star= c^{jk}_l x^l
\ee
do admit a generalisation of \eqn{der} according to 
\be
D_j f = k [x^j, f]_\star
\ee
with $k$ a suitable dimensionful constant, but the limit, $\Theta\rightarrow 0$, does not yield the correct commutative limit (see \cite{MVZ,  pvitale, GVW14}  for details and applications). A related  approach,  is to use twisted differential calculus for those NC algebras whose star product is defined in terms of a twist \cite{Vassilevich,chaichian, twist}. 

To summarise, for generic coordinate dependence of $\Theta$, where 
one needs to employ the general Kontsevich star product \cite{Kontsevich}, 
\be \label{Kon}
f\star g=f\cdot g+\frac{i}{2}\,\Theta^{ab}(x)\,\partial_a f\partial_b g+\dots\,,
\ee
   ordinary derivations violate  Leibniz rule,
\begin{equation*}
\partial_c(f\star g)=(\partial_c f)\star g+f\star(\partial_c g)+\frac{i}{2}\,\partial_c\Theta^{ab}(x)\,\partial_a f\partial_b g+\dots\,
\end{equation*}
whereas twisted or star derivations, although giving rise to a well defined differential calculus,  might not   reproduce the correct commutative limit.   The problem is not new and several attempts to its solution can be found in the literature. 

Other than identifying a differential calculus which be compatible with noncommutativity and yield back the correct commutative limit, we mentioned another problem which emerges in NCQFT, that is the UV/IR mixing, which certainly affects QFT with constant noncommutativity and may or may not affect coordinate dependent cases. Given the important role that NC field theory may play as an effective field theory implementing quantum gravity effects in some low energy regime \cite{Doplicher}, it is therefore worth to explore a novel approach which might help overcoming some of the  problems encountered so far.

Here we shall consider specifically pure gauge theories (no matter fields) and address the problem from a different perspective. Namely, we shall investigate how to modify the very definition of  gauge fields and gauge transformations in such a way that they be compatible with space-time noncommutativity and reproduce the correct commutative limit. We propose a novel strategy, which is inspired  by a recent approach to    gauge theories  \cite{BBKL, {kup-durham}}, which is  based on the conjecture that any well defined  gauge theory, including noncommutative and non-associative ones,  can be consistently constructed by bootstrapping some starting commutative gauge theory with a noncommutative (resp. non-associative) deformation in such a way to complete some $L_\infty$ algebra (see \cite{HZ17} for a physically oriented review of the role of $L_\infty$ algebras in field theory).   However the purpose of the paper is to show that it is possible to follow  a simpler, constructive approach,  which  can be autonomously understood, without recurring to the technical complexity of dealing with $L_\infty$ algebras. %without recurring to  as a powerful instrument to address the problem in the appropriate perspective .
% We mention here the method of the covariant coordinates formulated by Wess and collaborators in
%\cite{Madore:2000en}, where the partial derivative $\partial$ is substituted with the inner one, $D_a=c[x_a,\cdot]_\star$. For the associative star product the Leibniz rule for the star commutator, 
%\begin{equation*}
%D_a[f,g]_\star=[D_a f, g]_\star+[f,D_ag]_\star\,,
%\end{equation*}
% follows from the Jacobi identity, 
% \begin{equation*}
%[x_a,[f,g]_\star]_\star+[f,[g,x_a]_\star]_\star+[g,[x_a, f]_\star]_\star\equiv0\,.
%\end{equation*}
% However, since for, $\Theta\to0$, the star commutator vanishes, as well as the kinetic term in the action. So, in general, the commutative limit in this model is not well defined.
 
 Before proceeding further, an important remark is in order. Although the procedure is well defined for general   space-time non-commutativity, we shall work within a simplified scheme, which amounts to replace $\star$ commutators with Poisson brackets. Then, in order for the construction to be consistent, as long as the product is considered, in Eq. \eqn{Kon} only the zeroth order in the deformation parameter has to be  retained. Strictly speaking this means that space-time stays commutative, but it becomes a Poisson manifold with non-trivial Poisson bracket among position coordinates. Gauge parameters in turn, which are space-time functions, inherit such a  non-trivial Poisson structure.  It is therefore  natural  to require that they close under  Poisson brackets and it is a legitimate question to ask how gauge theories  have to be  modified in order to preserve gauge covariance. We shall see that there is no conceptual issue in generalising the whole  construction to a genuine non-commutative spacetime, although computationally more complicated.    

The paper is organised as follows. In Section \ref{ncga} the Poisson algebra of deformed gauge transformations is introduced and two guiding  examples are described, respectively with constant and Lie algebra type noncommutativity.  In Section \ref{ncfs} a recursive equation for the field strength definition is established, and solved order by order in the deformation parameter. Sections \ref{nccs} and \ref{ncym} contain respectively applications to Chern-Simons and Yang-Mills theories. In Section \ref{concl} we summarise our findings and discuss future perspectives. Finally, Appendix A contains a short review of gauge connections and field strength in derivation-based NC gauge theories.
 
\section{Non-commutative $U(1)$ gauge algebra}\label{ncga}

Let us consider   noncommutative space-time represented by the algebra $\mathcal{A}_\Theta$ with non-constant non-commutativity parameter $\Theta(x)$\footnote{ We use the symbol capital $\Theta$  for the NC tensor, $[x^i,x^j]=\Theta^{ij}(x)$, and lowercase  $\theta$ to indicate  a small, real parameter.}. We look for a deformed  theory of gauge fields which satisfies two main requirements: it is gauge invariant and reproduces in the commutative limit, $\Theta\to 0$, the standard gauge theory.  

For conventional   $U(1)$ gauge transformations,
$
\delta^0_f A=\partial f, 
$\,
gauge parameters close un  Abelian algebra, $[\delta^0_f,\delta^0_g] = 0$. For non-Abelian gauge theories where  gauge parameters are  valued in a non-Abelian Lie algebra , ${\mathbf f} = f_i \tau^i$, we have instead $\delta^0_{\mathbf f} A= \del {\mathbf f} -i  [A, {\mathbf f}] $ so that 
$$[\delta^0_{\mathbf f},\delta^0_{\mathbf g}] A= \del [{\mathbf f}, {\mathbf g}] -i  [A,[{\mathbf f},{\mathbf g} ]]= \delta^0_{ [{\mathbf f}, {\mathbf g}]}A.$$ 
Namely, the algebra of gauge parameters closes with respect to a non-Abelian Lie bracket. Noncommutative  $U(1)$ gauge theory, with gauge parameters now  belonging to $\mathcal{A}_\Theta$ behaves very much like non-Abelian  theories in many respects. Therefore  we shall require that the algebra of gauge parameters closes with respect to the star commutator, namely 
\begin{equation}
[\delta_f,\delta_g]A=\delta_{-i[f,g]_\star}A\,.\label{gas}
\end{equation}
However, if gauge connections  are defined as in appendix \ref{appa}, with gauge transformation
\be\label{ncgautr}
A'= A+ \del f -i [A, f]_\star
\ee
by composing two such transformations we get the result \eqn{gas} only if $\del$ is a derivation of the star commutator, which, as we have discussed in previous section, in general is not the case. 

Our aim in this section is somehow dual to what is usually done, namely, instead of looking for a deformed differential calculus, we shall deform the very definition of  gauge transformations, 
\be\label{gautra}
\delta^0_f A\to\delta_f A =\partial f+\dots
\ee
in such a way that \eqn{gas} be satisfied.

As  already noted, we shall work in the  slowly varying, but not necessarily small fields. In such a case we discard  higher derivatives terms in the star commutator and take, 
\begin{equation}
-i[f,g]_\star\approx \{f,g\}=\Theta^{ab}(x) \,\partial_a f\,\partial_b g\,.\label{1}
\end{equation}
In the approximation which we have chosen Eq. \eqn{gas} becomes
\begin{equation}
[\delta_f,\delta_g]A=\delta_{\{f,g\}}A\,.\label{ga}
\end{equation}
and we look for  gauge transformations in the form  \eqn{gautra} which be compatible with the latter. 

A remark is here in order. In the chosen approximation, space-time is still commutative, namely the product between fields is the usual point-wise product, but its geometry is deformed, because it acquires a non-trivial Poisson bracket. Therefore one should rather talk about commutative field theory on Poisson-deformed space-time. Once such a distinction made, in the following we shall refer to the latter as noncommutative space time without any further specification, unless otherwise stated.  

A solution to this problem has been proposed in \cite{BBKL,Kup27} in terms of  field dependent gauge transformations, in the form 
%\textcolor{magenta}{If in the cited references you made the explicit assumption that theta be linear in x, then it would be good to stress here that we have verified that, the result found  for the linear assumption, has now be demonstrated to be valid in general. Otherwise it seems that everything was already done.} \textcolor{blue}{Conceptually there is nothing new in this section, all (including the $\Theta$ in general form) was already done in \cite{Kup27}. The only new thing here is a more convenient notation, $\gamma^k_a(A)=\delta^k_a+\Gamma^k_a(A)$. The essentially new results in this paper are included in the Sections 3 and 5.}
\begin{equation}
\delta_f A_a=\gamma_a^k(A)\,\partial_k f+ \{A_a,f\}\,,\label{gt}
\end{equation}
Indeed, it may be verified that the latter close the algebra (\ref{ga}) if the matrix $\gamma(A)_a^k$ satisfies the equation,\footnote{The convention used here is: the partial derivative with the upper index is the derivation with respect to the field, $\partial^b_A=\partial/\partial A_b$, while the partial derivative with the lower index is a derivation with respect to coordinate, $\partial_m=\partial/\partial x^m$.}
\begin{equation}
\gamma^l_b\,\partial^b_A\gamma_a^k-\gamma^k_b\,\partial^b_A\gamma^l_a+\Theta^{lm}\,\partial_m\gamma_a^k-\Theta^{km}\,\partial_m\gamma_a^l-\gamma^m_a\,\partial_m\Theta^{lk}=0\,,\label{eq1}
\end{equation}
where we set, $ \gamma^{k(0)}_a=\delta^k_a$, to ensure the correct commutative limit.
For arbitrary non-commutativity parameter $\Theta^{kl}(x)$ Eq.  (\ref{eq1}) was solved   in the form of a perturbative series \cite{Kup27},
\begin{eqnarray}
\gamma^k_a(A)=\sum_{n=0}^\infty\,\gamma^{k(n)}_a&=&\delta^k_a-\frac12\, \partial_a \Theta^{kb} A_b\label{gps}\\
&&-\frac{1}{12}\left(2\,\Theta^{cm}\partial_a\partial_m\Theta^{bk}+\partial_a\Theta^{bm}\partial_m\Theta^{kc}\right)A_bA_c+{\cal O}(\Theta^3)\,.\notag
\end{eqnarray}
Note that the order of  each term $\gamma^{k(n)}_a$ in the gauge fields $A$ coincides with the order of this term in the deformation parameter $\Theta$. We also stress here that the Ansatz in Eq. (\ref{gt}) %for the NC deformation of the gauge transformations 
 takes into account only the leading order contribution in derivatives $\partial f$ and $\partial A$. However  all orders in $\Theta$ are included, this being necessary to close the algebra (\ref{ga}). In this sense Eq. (\ref{gt}) is exact.

For some specific choices of  non-commutativity, one may also discuss the convergence of the series (\ref{gps}) and exhibit a closed expression  for the gauge transformation (\ref{gt}). Here we discuss two particular cases.  

\subsubsection*{Canonical non-commutativity} 

Canonical non-commutativity corresponds to  constant NC parameter $\Theta^{kl}$. Since, $\partial_m\Theta^{kl}=0$, the constant solution $ \gamma^{k}_a=\delta^k_a$ solves Eq.  (\ref{eq1}), yielding the gauge transformations,
\begin{equation}
\delta_f A_a=\partial_a f+ \{A_a,f\}\,.\label{gtc}
\end{equation}
Let us notice here that $\gamma^{k}_a=\delta^k_a$ is also a solution for the fully noncommutative case where we replace Poisson brackets by $\star$-commutators. Eq. \eqn{gtc} becomes  $\delta_f A_a=\partial_a f-i [A_a,f]_\star\,.$, which coincides with the standard definition of NC gauge transformation \eqn{ncgautr} and Eq. \eqn{gas} is satisfied.
\subsubsection*{Lie algebra noncommutativity: $\R^3_\theta$ } 

The three dimensional rotationally invariant non-commutative space,  $\R^3_\theta$, \cite{Hammou:2001cc,GraciaBondia:2001ct,pvitale,Galikova:2013zca,Kupriyanov:2012nb,Kupriyanov:2015uxa} corresponds in the approximation we have chosen,  to the $su(2)$-like Poisson algebra,
\begin{equation}
\{x^k,x^l\}=2\,\theta\,{\varepsilon^{kl}}_m\, x^m\,,\label{su2}
\end{equation}
where the real number $\theta$ is a small parameter and $\varepsilon^{klm}$ is the Levi-Civita symbol. The factor of 2 is just a matter of convenience. In this case the solution of  equation (\ref{eq1}) reads \cite{kup-durham}
\begin{equation}
\gamma^k_a(A)=\left[1+\theta^2A^2\chi\left(\theta^2A^2\right)\right]\delta^k_a-\theta^2\chi\left(\theta^2A^2\right)A_aA^k-\varepsilon_a{}^{kl}A_l\,,\label{gammasu2}
\end{equation}
where
\eq{
\chi(t)=\frac1t\,\left(\sqrt{t}\cot\sqrt{t}-1\right)\,,\qquad \chi(0)=-\frac13\,.
} 
We use the Kronecker delta to raise and lower  indices, %e.g., $\varepsilon_a{}^{kl}=\delta_{am}\varepsilon^{mkl}$, 
and  summation under the repeated indices is understood, $A^2=A_mA^m$.

The corresponding non-commutative deformation of  Abelian gauge transformations reads \cite{kup-durham},
\eq{\label{gt1}
 \delta_{f}  A_a=\partial_af+\{A_a,f\}+\theta\,\varepsilon_a{}^{kl}A_k\partial_lf+\theta^2\,\left(\partial_af A^2-\partial_kfA^kA_a\right)\chi\left(\theta^2A^2\right)\,.
}
The latter may be verified to  close the algebra (\ref{ga}) and to reproduce the correct commutative limit $\theta\to0$. 

This result is essentially different from what one would get in standard approaches.  See  for example \cite{GVW14}, where the infinitesimal transformation of the gauge potential for Lie $\mathfrak{su}(2)$-type noncommutativity reads
$ 
\delta_f A_a= D_a f +i[f, A_a]_*
$
with $D_a = -\frac{i}{\theta}[x_a,\cdot]_* \stackrel{\theta\rightarrow 0}{\rightarrow} \epsilon_{abc} x_b \del_c$. Lie algebra type $\star$-commutators do not converge to usual derivations in the commutative limit and the whole gauge theory behaves quite differently from the commutative analogue (see \cite{GVW14}  for related duscussion).

\section{Non-commutative field strength}\label{ncfs}

In previous section the $U(1)$ gauge potential has been introduced as a vector-valued element of the NC algebra $\mathcal{A}_\Theta$, $\{A_a\}$, $a=1,\ldots  {\rm dim} \mathcal{A}_\Theta$, whose gauge transformation \eqn{gt} was fixed by the request that it be compatible with the closure of the algebra of gauge parameters \eqn{ga}.
Similarly, we  look here  for a deformation of the $U(1)$ field strength,
\begin{equation}
{\cal F}_{ab}=\partial_aA_b-\partial_bA_a+{\cal O}(\Theta)\,,\label{F0}
\end{equation}
which  be covariant under  gauge transformations (\ref{gt}), namely satisfying
\begin{equation}
\delta_f {\cal F}_{ab}=\{{\cal F}_{ab},f\}\,,\label{gcc}
\end{equation}
where, $\delta_f {\cal F}_{ab}:={\cal F}_{ab}(A+\delta_fA)-{\cal F}_{ab}(A)\,.$ In three space-time dimensions such a field  was constructed in \cite{Kup27}.  

In this section we address the general  $n$-dimensional case. Following \cite{Kup27} we are looking for a  solution of Eq.  (\ref{gcc}) in the form,
\begin{equation}
{\cal F}_{ab}=P_{ab}{}^{cd}\left(A\right)\,\partial_c A_d+R_{ab}{}^{cd}\left(A\right)\,\left\{A_c,A_d\right\}\,,\label{4}
\end{equation}
where we choose
\begin{equation}\label{PRexp}
P_{ab}{}^{cd}\left(A\right)=\delta_a^{c}\delta_b^{d}-\delta_a^{d}\delta_b^{c}+{\cal O}(\Theta)\,,\,\, R_{ab}{}^{cd}\left(A\right)=\frac12\left(\delta_a^{c}\delta_b^{d}-\delta_a^{d}\delta_b^{c}\right)+{\cal O}(\Theta)\,,
\end{equation}
to match (\ref{F0}). By construction, $R_{ab}{}^{cd}\left(A\right)$ should be antisymmetric in  upper indices since it is contracted with the Poisson bracket $\left\{A_c,A_d\right\}$.

Eq.(\ref{gcc}), upon replacing the   Ansatz (\ref{4}), becomes after simplification,
 \begin{eqnarray}
  &&\left[\gamma^k_l\,\partial^l_A P_{ab}{}^{cd}+\Theta^{kl}\,\partial_lP_{ab}{}^{cd}+P_{ab}{}^{cl}\,\partial^d_A\gamma^k_l+P_{ab}{}^{ld}\,\partial_l\Theta^{ck}+2\,R_{ab}{}^{ld}\,\partial_m\gamma^k_l\,\Theta^{mc}\right]\,\partial_cA_d\,\partial_kf+\notag\\
&&P_{ab}{}^{cd}\,\gamma^k_d\,\partial_c\partial_kf+\left[P_{ab}{}^{cd}-2\,\gamma^c_lR_{ab}{}^{ld}\right]\,\{A_d,\partial_cf\}+\notag\\
&&\left[\gamma^k_l\,\partial^l_A R_{ab}{}^{cd}+\Theta^{kl}\,\partial_lR_{ab}{}^{cd}+R_{ab}{}^{cl}\,\partial^d_A\gamma^k_l+R_{ab}{}^{ld}\,\partial^c_A\gamma^k_l\right]\left\{A_c,A_d\right\}\,\partial_kf=0\,.\label{cond}
\end{eqnarray}
The latter should hold for any gauge parameter $f$ and any gauge field $A_a$. Thus, Eq.  (\ref{gcc}) yields four separate equations for  the coefficient functions $P_{ab}{}^{cd}$ and $R_{ab}{}^{cd}$. The first equation involves $P_{ab}{}^{cd}$ and $R_{ab}{}^{cd}$,
\begin{eqnarray}\label{e8a1}
\gamma^k_l\,\partial^l_A P_{ab}{}^{cd}+\Theta^{kl}\,\partial_lP_{ab}{}^{cd}+P_{ab}{}^{cl}\,\partial^d_A\gamma^k_l+P_{ab}{}^{ld}\,\partial_l\Theta^{ck}+2\,R_{ab}{}^{ld}\,\partial_m\gamma^k_l\,\Theta^{mc}=0\,.
\end{eqnarray}
The second one,
\begin{equation}
P_{ab}{}^{[cd}\,\gamma^{k]}_d=0\,,\label{e8a2}
\end{equation}
is an algebraic relation on the coefficient $P_{ab}{}^{cd}$. The third equation relates $P_{ab}{}^{cd}$ and $R_{ab}{}^{cd}$,
\begin{equation}
P_{ab}{}^{cd}=2\,\gamma^c_lR_{ab}{}^{ld}\,\label{PR}
\end{equation}
and the last one is an  equation for  $R_{ab}{}^{cd}$ reading,
\begin{equation}
\gamma^k_l\,\partial^l_A R_{ab}{}^{cd}+\Theta^{kl}\,\partial_lR_{ab}{}^{cd}+R_{ab}{}^{cl}\,\partial^d_A\gamma^k_l+R_{ab}{}^{ld}\,\partial^c_A\gamma^k_l=0\,.\label{eqR}
\end{equation}
To start with, we look for a perturbative solution  in $\Theta$ of  equation (\ref{eqR}) using  the expression for $\gamma^k_l$  found  previously, Eq. \eqn{gps}. On using the second of Eqs. \eqn{PRexp} up to first order in $\Theta$ and observing that,  at  first order $\gamma^{k(1)}_l=-\partial_l \Theta^{kb} A_b/2$, one obtains from (\ref{eqR}),
\begin{equation}
R_{ab}{}^{cd(1)}=\frac14\left(\delta^c_a\,\partial_b\Theta^{kd}-\delta^d_a\,\partial_b\Theta^{kc}-\delta^c_b\,\partial_a\Theta^{kd}+\delta^d_b\,\partial_a\Theta^{kc}\right)A_k\,.\label{R1}
\end{equation}
As for the second order, substituting the latter  back into (\ref{eqR}) and using the expression for  $\gamma^{k(2)}_l$ given in (\ref{gps}) one finds, 
\begin{eqnarray}
R_{ab}{}^{cd(2)}&=&\left(\frac{1}{12}\delta^c_a\,\Theta^{nm}\,\partial_b\partial_m\Theta^{kd}-\frac{1}{12}\delta^d_a\,\Theta^{nm}\,\partial_b\partial_m\Theta^{kc}\right.\label{R2}\\&&-\frac{1}{12}\delta^c_b\,\Theta^{nm}\,\partial_a\partial_m\Theta^{kd}+\frac{1}{12}\delta^d_b\,\Theta^{nm}\,\partial_a\partial_m\Theta^{kc}\notag\\
&&+\frac{1}{12}\delta^c_a\,\partial_b\Theta^{nm}\,\partial_m\Theta^{kd}-\frac{1}{12}\delta^d_a\,\partial_b\Theta^{nm}\,\partial_m\Theta^{kc}\notag\\&&-\frac{1}{12}\delta^c_b\,\partial_a\Theta^{nm}\,\partial_m\Theta^{kd}+\frac{1}{12}\delta^d_b\,\partial_a\Theta^{nm}\,\partial_m\Theta^{kc}\notag\\
&&+\left.\frac{1}{8}\partial_a\Theta^{kc}\,\partial_b\Theta^{nd}-\frac{1}{8}\partial_a\Theta^{nd}\,\partial_b\Theta^{kc}\right)\,A_kA_n\,.\notag
\end{eqnarray}
The process can be thus iterated to all orders in $\Theta$. 

It is remarkable that since Eq.  (\ref{PR}) expresses the coefficient functions $P_{ab}{}^{cd}$ in terms of $R_{ab}{}^{cd}$ and $\gamma^c_l$, Eqs.  (\ref{e8a1}) and (\ref{e8a2}) become consistency conditions for the solution of Eq.  (\ref{gcc}). One may check that (\ref{e8a1}) holds as a consequence of (\ref{PR}), (\ref{eqR}) and (\ref{eq1}), while (\ref{e8a2}) is satisfied as a consequence of (\ref{PR}) and the antisymmetry of  $R_{ab}{}^{cd}$.

Let us notice that the result we have found for the strength field $\mathcal{F}$ is valid in any dimension, whereas  previous result in \cite{Kup27} was specific of three dimensions. Moreover, the latter   was only valid for linear $\Theta$, while now we have considered a general dependence in $x$.  In the linear case  , $\Theta^{kl}(x)=c^{kl}_mx^m$, the coefficient functions $\gamma^k_a(A)$, $P_{ab}{}^{cd}(A)$ and $R_{ab}{}^{cd}(A)$ are only  functions of the gauge field $A$ and do not depend  explicitly on coordinates. For general $\Theta^{kl}(x)$ they  may have explicit   $x$ dependence.
%, $\gamma^k_a(x,A)$, $P_{ab}{}^{cd}(x,A)$ and $R_{ab}{}^{cd}(x,A)$. 
This in turn  produces additional contributions of the form $\Theta^{kl}\,\partial_lP_{ab}{}^{cd}$, $2\,R_{ab}{}^{ld}\,\partial_m\gamma^k_l\,\Theta^{mc}$ and $\Theta^{kl}\,\partial_lR_{ab}{}^{cd}$  which have been included in Eqs.  (\ref{cond})-\ref{eqR}).

\subsubsection*{Canonical non-commutativity} 

Since in this case, $\gamma^k_l=\delta^k_l$, one finds from  Eqs. (\ref{PR}), (\ref{eqR}) ,
\begin{equation}\label{RPc}
 R_{ab}{}^{cd}\left(A\right)=\frac12\left(\delta_a^{c}\delta_b^{d}-\delta_a^{d}\delta_b^{c}\right),\qquad\mbox{and}\qquad P_{ab}{}^{cd}\left(A\right)=\delta_a^{c}\delta_b^{d}-\delta_a^{d}\delta_b^{c}\,,
\end{equation}
which results in,
\begin{equation}
{\cal F}^{can}_{ab}=\partial_aA_b-\partial_bA_a+\{A_a,A_b\}\,.\label{Fcc}
\end{equation}
Similarly to the result we found for  the gauge potential, if we repeat  the procedure just described  for  a fully canonical non-commutative theory, with Poisson brackets replaced by star commutators, one  obtains \cite{BBKL} 
\begin{equation}
{\cal F}^{NC}_{ab}=\partial_aA_b-\partial_bA_a-i[A_a,A_b]_\star\,.\label{Fc1}
\end{equation}
which is consistent   with the standard definition of NC  field strength \eqn{fmunu}.

\subsubsection*{Lie algebra noncommutativity: $\R^3_\theta$ } 
For $\mathfrak{su}(2)$-like noncommutativity, in the slowly varying fields approximation,  the matrix $\gamma^k_l(A)$ was determined in (\ref{gammasu2}). The solution of  equation (\ref{eqR}) reads,
\begin{eqnarray}
 R_{ab}{}^{cd}\left(A\right)&=&\frac12\left(\delta_a^{c}\delta_b^{d}-\delta_a^{d}\delta_b^{c}\right)\,\lambda\left(\theta^2A^2\right)+\label{Rsu2}\\
 &&\frac{\theta}{2}\left(\varepsilon_{ab}{}^cA^d-\varepsilon_{ab}{}^dA^c\right)\,\lambda\left(\theta^2A^2\right)+\notag\\
 &&\frac{\theta^2}{2}\left(\delta^c_aA_bA^d-\delta^c_bA_aA^d-\delta^d_aA_bA^c+\delta^d_bA_aA^c\right)\,\lambda^\prime\left(\theta^2A^2\right)\,,\notag
\end{eqnarray}
where
\begin{equation}
\lambda(t)=\left(\frac{\sin\sqrt{t}}{\sqrt{t}}\right)^2\,
\end{equation}
and $\lambda'$ indicates its derivative. 
The function $\lambda(t)$ satisfies the equation, $\lambda^\prime=\chi \lambda$, with initial condition, $\lambda(0)=1$.
Being  in $3d$ any totally antisymmetric tensor of rank four vanishes. In particular,
\begin{equation}
\varepsilon^{abc}\,A^e-\varepsilon^{bce}\,A^a+\varepsilon^{cea}\,A^b-\varepsilon^{eab}\,A^c=0\,.
\end{equation}
By taking into account the above relation and its consequences in Eq. (\ref{PR}) we represent the coefficient $P_{ab}{}^{cd}$ in a more convenient form,
\begin{eqnarray}
 P_{ab}{}^{cd}\left(A\right)&=&\left(\delta_a^{c}\delta_b^{d}-\delta_a^{d}\delta_b^{c}\right)\phi\left(\theta^2A^2\right)+2\,\theta\,\varepsilon_{ab}{}^cA^d\,\phi\left(\theta^2A^2\right)\label{Psu2}\\
 &&-\theta\,\varepsilon_{abm}A^m\delta^{cd}\,\lambda\left(\theta^2A^2\right)-\theta\,\varepsilon_{ab}{}^dA^c\,\lambda\left(\theta^2A^2\right)\notag\\
 &&+\theta^2\left(\delta^c_aA_bA^d-\delta^c_bA_aA^d\right)\left[\chi \phi-\lambda\right]\left(\theta^2A^2\right)\notag\\&&-\theta^3\varepsilon_{abm}A^mA^cA^d\,\lambda^\prime\left(\theta^2A^2\right)\,,\notag
\end{eqnarray}
with,
\begin{equation}
\phi(t)=(1+t\chi(t))\,\lambda(t)=\frac{\sin\sqrt{t}\cos\sqrt{t}}{\sqrt{t}}\,.
\end{equation}
It is remarkable that there are only two independent functions $\chi(t)$ and $\lambda(t)$ which determine the whole construction.

In this case we can simplify the expression for the field strength. One may check that,
\begin{equation}
\partial^l_AR_{ab}{}^{cd}+\mbox{cycl.}(cdl)=0\,.\label{dR}
\end{equation}
Consequently,
\begin{equation}
R_{ab}{}^{cd}=\partial^c_A\,\rho_{ab}{}^d-\partial^d_A\,\rho_{ab}{}^c\,,
\end{equation}
where
\begin{equation}
\rho_{ab}{}^c=\frac14\left(\delta_b^cA_a-\delta_a^cA_b\right)\,\lambda\left(\theta^2A^2\right)-\frac{1}{4\theta}\varepsilon_{ab}{}^c\,\Lambda\left(\theta^2A^2\right)\,,
\end{equation}
with, $\Lambda^\prime(t)=\lambda(t)$. The same procedure  can be applied for  $ P_{ab}{}^{cd}$. Since,
\begin{equation}
 \partial^l_AP_{ab}{}^{cd}=\partial^d_AP_{ab}{}^{cl}\,,\label{dP}
\end{equation}
we may represent it as,
\begin{equation}
P_{ab}{}^{cd}=\partial^d_A\,\pi_{ab}{}^c\,,
\end{equation}
where
\begin{equation}
\pi_{ab}{}^c=\left(\delta_a^cA_b-\delta_b^cA_a\right)\,\phi\left(\theta^2A^2\right)-\theta\,\varepsilon_{abm}A^mA^c\,\lambda\left(\theta^2A^2\right)+\frac{1}{\theta}\varepsilon_{ab}{}^c\,\Phi\left(\theta^2A^2\right)\,,
\end{equation}
with, $\Phi(t)=\int\phi(t)dt=-\cos(2\sqrt{t})/2$. Then the expression for the non-commutative field strength becomes,
\begin{equation}
{\cal F}^{su(2)}_{ab}=\partial_c\,\pi_{ab}{}^c+2\,\{\rho_{ab}{}^c,A_c\}\,.\label{Fsu2}
\end{equation}
In the standard approach  the field strength is defined as in \eqn{courgene}. Then the Bianchi identity is satisfied by definition. See for example \cite{GVW14} for a comparison in case of $\mathfrak{su}(2)$-like noncommutativity. 

Within the present approach Bianchi identity is not automatically built-in because the field strength is not defined as the curvature of a connection. However one may still ask whether the non-commutative field strength (\ref{Fsu2}) satisfies some  deformed Bianchi identity. We leave it as an open problem.

\section{Noncommutative Chern-Simons model}   \label{nccs}

Noncommutative deformation of Chern-Simons (CS) theory was constructed in \cite{Kup27}. In this section for  completeness and for the convenience of the reader we recollect the main findings of \cite{Kup27}. Just like in the standard commutative case,  non-commutative Chern-Simons equations are obtained by requiring that the NC Field strength should vanish everywhere. Since we are in three dimensions we may set,
\begin{equation}
{\mathcal F}^a(A):=\frac12\varepsilon^{abc}{\cal F}_{bc}=P^{abc}\left(A\right)\,\partial_b A_c+R^{abc}\left(A\right)\,\left\{A_b,A_c\right\}=0\,.\label{7}
\end{equation}
With respect to the comment made at the end of last section, let us notice that  this definition of the field strength looks like a deformation of the covariant derivative, through the coefficient functions 
$P^{abc}$ and $R^{abc}$.  Eq. \eqn{7} satisfies the following two requirements. 
It transforms covariantly under the NC gauge transformations (\ref{gt}), $ \delta_{f}  {\mathcal F}^a = \{{\mathcal F}^a,f\}\,,$ and reproduces in the commutative limit, $\Theta\to 0$, the standard Abelian CS equation of motion, i.e., $\lim_{\Theta\to0}\,{\mathcal F}^a(A)=\varepsilon^{abc}\partial_bA_c$. These two properties are exactly what  we expect from a suitable  noncommutative deformation of  Chern-Simons theory. 

It is important to stress that the noncommutative CS equations (\ref{7}) are non-Lagrangian. Indeed, they do not satisfy the criterium of commutation of second variational derivatives,
\begin{equation*}
\frac{\delta {\cal F}^a}{\delta A^b}\neq\frac{ \delta{\cal F}^b}{\delta A^a}\,.
\end{equation*}
This is a main difference between our proposal and  previous approaches. It is not based on a deformation of the commutative action as a consequence of  a modification of the geometric structures involved, but on the request of gauge covariance  and the correct commutative limit of the corresponding field equations. The dynamics  that is  obtained in such a way may not admit the existence of an action principle, as it is the case for CS equations (\ref{7}).
 
\section{Non-commutative Yang-Mills theory}\label{ncym}

Having  defined the NC field strength as in (\ref{4}), differently form CS dynamics,   it is possible for $U(1)$ Yang-Mills theory to     introduce a non-commutative deformation  by means of an action principle. 

On defining the non-commutative Yang-Mills Lagrangian as,
\begin{equation}
{\mathcal L}=-\frac14  {\cal F}_{ab} {\cal F}^{ab}\,\label{LYM}
\end{equation}
it is possible to verify that it  transforms covariantly under the NC gauge transformations (\ref{gt}),
$
 \delta_{f}  {\cal L} = \{f,{\cal L}\}\,.
$
 Consequently the corresponding action, $S=\int {\mathcal L}$, is gauge invariant, $ \delta_{f} S\equiv0$. By Noether's second theorem the gauge invariance of the action functional implies the existence of non-trivial differential relations (Noether identities) among the corresponding Euler-Lagrange equations. See, e.g., \cite{Ciric:2020hfj} for the derivation of the Noether identities within the L$_\infty$-formalism. 

\subsubsection*{Canonical noncommutativity }
For canonical noncommutativity it is particularly easy to derive the equations of motion which yield  the  noncommutative analogue of Maxwell equations. Taking into account Eq. (\ref{RPc}) for  the coefficient functions $R_{ab}{}^{cd}$ and $P_{ab}{}^{cd}$ which correspond to constant $\Theta$, we get from the Lagrangian (\ref{LYM}) Euler Lagrange equations  in the form:
\begin{equation}
\partial_a\,{\cal F}_{can}^{ad}+\{A_a, {\cal F}_{can}^{ad}\}=0\,
\end{equation}
where ${\cal F}^{can}_{ab}$ was found in Eq. (\ref{Fcc}) to be 
\begin{equation*}
{\cal F}^{can}_{ab}=\partial_aA_b-\partial_bA_a+\{A_a,A_b\}\,.
\end{equation*}
They correspond to NC Maxwell equations of the standard approach, when the star commutator is replaced with the Poisson bracket, and reproduce  the correct commutative limit as expected. 
 \subsubsection*{Lie algebra noncommutativity: $\R^3_\theta$ }
 Taking into account that for the $\mathfrak{su}(2)$-like non-commutativity the coefficient functions $R_{ab}{}^{cd}$ and $P_{ab}{}^{cd}$ satisfy Eqs.  (\ref{dR}) and (\ref{dP}), one finds for the Euler-Lagrange equations corresponding to (\ref{LYM}),
\begin{equation}
 {\cal E}^d(A):={\cal D}_{ab}{}^{d}\,{\cal F}^{ab}=0\,,
\end{equation}
where the field strength is defined in Eq. \eqn{Fsu2} and 
\begin{equation}\label{eomsu2}
{\cal D}_{ab}{}^{d}\,{\cal F}^{ab}=\frac12 P_{ab}{}^{cd}(A)\,\partial_c\, {\cal F}^{ab}-R_{ab}{}^{cd}(A)\,\{A_c, {\cal F}^{ab}\}\,.
\end{equation}
By  construction these equations are gauge-covariant and reproduce the $U(1)$ Yang-Mills equations in the commutative limit. 

It is interesting to notice that Eq. \eqn{eomsu2} acquires the form of a deformed covariant derivative, with the coefficient functions $R_{ab}{}^{cd}$ and $P_{ab}{}^{cd}$ taking care of the deformation.  We  plan to come back to this issue elsewhere.

\section{Conclusions}\label{concl}
In the paper we have proposed a novel approach to $U(1)$ noncommutative gauge theory, which is based on the request that the commutative limit be retrieved for $\Theta\rightarrow 0$ and  the dynamics of pure gauge fields be gauge covariant. This is achieved by constructively defining the gauge potential and the field strength through recursive equations which may be solved order by order in the NC parameter, and, generalising   previous derivations  \cite{BBKL, kup-durham, Kup27},  are valid in any space-time dimension and for generic dependence of $\Theta$ on space-time coordinates. As for the examples considered, especially interesting is the Lie-algebra type noncommutativity. We are presently investigating another instance of such a family, which is the so-called k-Minkowski spacetime \cite{KKV20}. 

Two immediate research questions which would be interesting to investigate  are the following. First, prior to any quantum field theory application, one should solve the classical equations of motion and check fundamental problems such as  the propagation of light in such a deformed  space-time. Secondly, one should address the problem of coupling gauge fields with matter fields. Indeed, since the gauge potential is not introduced as a connection one-form, the  notion of  covariant derivative in such a theory is not automatic. A related problem is the possibility of reformulating   the above defined field strength as a consistent deformation  of the notion of curvature of the gauge potential. We plan to come back to these issues in the near future.
 
\subsection*{Acknowledgments}

V.G.K. thanks the University of Naples for hospitality where the present work was initiated. 

\begin{appendix}
\section{Gauge connection}\label{appa}%

 %
%\begin{eqnarray*}
%d\omega(X_1,..., X_{p+1})& :=& \sum_{i=1}^{p+1} (-1)^{i+1} X_i \omega( X_1,..\vee_i.., X_{p+1}) \\
%&+&  \sum_{1\leq i < j \leq p+1} (-1)^{i+j} \omega( [X_i, X_j],..\vee_i..\vee_j.., X_{p+1}) \label{eq:koszul},
%\end{eqnarray*}
%with $ X_j\in \DER((\mathcal{A})$, ${\Omega}^0_{\DER}(\mathcal{A})=\mathcal{A}$ ($\vee_i$ means that the argument $i$ is omitted). \par 

A natural noncommutative extension of the notion of connection is  introduced in \cite{dbv}, where one replaces complex  vector bundles of physical fields over space-time, %with fibre $\C^n$,   
with   right-modules, $\modul$ over  noncommutative space-time, namely the non-commutative algebra $\mathcal{A}$.  Generalising the standard definition which is proper  of geometric approaches to gauge theory,  a connection on $\modul$ can be conveniently defined %\cite{dbv1, dbv2, Wallet:2008bq} 
by a linear map ${\nabla} :  \DER(\mathcal{A})\times  \modul \rightarrow \modul$ satisfying
\begin{equation}
{\nabla}_X (m f) = mX( f) + {\nabla}_X (m) f,\ {\nabla}_{c X}( m) = c{\nabla}_X (m),\ 
{\nabla}_{X + Y} (m) = {\nabla}_X (m) + {\nabla}_Y (m) \label{connect}
\end{equation}
for any $X,Y \in \DER(\mathcal{A})$, $f \in \mathcal{A}$, $m \in \modul$, $c \in \caZ(\mathcal{A})$, the center of the algebra.  Hermitian connections  satisfy  for any real derivation $X \in \DER(\mathcal{A})$
\begin{equation}
 X(h(m_1,m_2))=h(\nabla_X(m_1),m_2)+h(m_1,\nabla_X(m_2)), \forall m_1,m_2\in\modul,\label{hermitconnect}
\end{equation}
where $h:\modul\otimes\modul\to\mathcal{A}$ denotes a Hermitian structure on $\mathcal{A}$. The curvature is the linear map $F(X, Y) : \modul \rightarrow \modul$ defined by
\begin{equation}
 F(X, Y) m = [ {\nabla}_X,{\nabla}_Y ] m - {\nabla}_{[X, Y]}m,\ \forall X, Y \in \DER(\mathcal{A})\label{courgene}.
\end{equation}

The group of gauge transformations   of $\modul$, ${\cal{U}}(\modul)$, is   defined  as the group of automorphisms of $\modul$ compatible both with the structure of right $\mathcal{A}$-module and the Hermitian structure, i.e 
\be
g(mf)=g(m)f,\;\;\; h(g(m_1),g(m_2))=h(m_1,m_2)\;\;\; \forall g\in{\cal{U}}(\modul), \;\;\; \forall m_1,m_2\in\modul \label{hermcond}
\ee
%This definition is the natural algebraic counterpart of Def. \ref{gaudef}. 
For any $g\in{\cal{U}}(\modul)$ we have 
\begin{eqnarray}
{\nabla}^g_X&:&\modul\to\modul,\ {\nabla}^g_X = g^{-1}\circ {\nabla}_X \circ g\label{gaugeconnect}\\
F(X,Y)^g&:&\modul\to\modul,\ F(X,Y)^g=g^{-1}\circ F(X,Y) \circ g\label{gaugecurv}.
\end{eqnarray}
For $U(1)$  gauge theory,  where the  relevant vector bundle   is a complex line bundle, the corresponding NC generalisation  is  a one-dimensional ${\mathcal{A}}$-module $\modul=\C \otimes\mathcal{A}$. As Hermitian structure one  chooses $h(f_1,f_2)=f_1^\dag f_2$ and takes real derivations. Then a Hermitian connection is entirely determined  by its action on the one-dimensional basis of the module, $\nabla_X({\mathbf{1}})$. We have  $ \nabla_X(f)= \nabla_X({\mathbf{1}}) f + X(f)$,with  $\nabla_X({\mathbf{1}})^\dag=\nabla_X({\mathbf{1}})$. This defines in turn the gauge connection 1-form,  $A$,  by  means of 
\begin{equation}
A:X\to A(X):=i \nabla_X({\mathbf{1}}), \;\; \forall X\in\DER(\mathcal{A})
\end{equation}
From the compatibility condition with the Hermitian structure, Eq. \eqn{hermcond}, one obtains  that  gauge transformations are  the group of unitary elements of the noncommutative algebra $\mathcal{A}$. Indeed, on using $g(f)= g( \mathbf{1}  f)= g( \mathbf{1})\star f $ and imposing compatibility, one obtains 
$ h(g(f_1), g(f_2))= h(f_1,f_2)$ which implies $ g(\mathbf{1})^\dag\star  g(\mathbf{1}) =\mathbf{1}$.   
We pose $g(\mathbf{1}) \equiv g \in \mathcal{U} (\mathcal{A}) $
  the group of unitary elements of  the NC algebra $\mathcal{A}$, acting multiplicatively on  $\mathcal{A}$ from the left. 
  
  To  give an explicit example of the whole construction, let us consider the $2$-dimensional Moyal plane, $\mathcal{A}= \R^{2}_\theta$,  with constant noncommutative parameter, $\theta$.  The latter is referred to as canonical noncommutativity in the paper. The algebra of derivations is in this case the Abelian algebra generated by derivatives $\del_\mu$. %= -i ${\theta^{-1}}_{\mu\nu}[x^\nu, \dot]_\star$
   From   Eqs. \eqref{gaugeconnect}, \eqref{gaugecurv} we obtain
   \be\label{fmunu}
  F_{\mu\nu}= F(\del_\mu, \del_\nu)=  \del_\mu A_\nu-\del_\nu A_\mu-i [A_\mu, A_\nu]_\star
  \ee 
 with $ A_\mu= i \nabla_\mu (\mathbf{1}), $  and, to make contact with usual  notation,  we have rescaled $F$  by a factor of $i$. 
 The unitary gauge group $\mathcal{U}(\R^2_\theta)$ acts as $\nabla_\mu^g = g^\dag \circ \nabla_\mu \circ g,$ yielding 
\begin{equation}
A_\mu^g=g\star A_\mu\star g^\dag-i \partial_\mu g \star g^\dag,\ F_{\mu\nu}^g=g\star F_{\mu\nu}\star g^\dag,\;\;\;\;\; \forall g\in{\cal{U}}(\mathcal{A})\label{gaugetrans}
\end{equation}
 
 Being  unitary elements of $\mathcal{A}$ gauge transformations may be written as star exponentials
  \begin{equation}
g[f]=  \exp_{\star}\left(i f\right),  \label{gtrans}
\end{equation}
and  the star exponential  is by definition 
\begin{equation}
\exp_{\star}(i f) \equiv \sum_{n=0}^{\infty} \frac{(i)^n}{n!}\underbrace{%
f\star...\star f}_{\mbox{$n$ times}}
\end{equation}
where the gauge parameters $f$ are  functions of $x\in \R^2_\theta$. 
Hence, 
for the Moyal plane we  get    infinitesimal gauge transformations in the form
\be\label{infgat}
\delta A_\mu=  \del_\mu f +i [f, A_\mu]_\star \;\;\; \delta F_{\mu\nu}=  i [f, F_{\mu\nu}]_\star.
\ee
%In the next section we shall study the infinitesimal form of Eq. \eqn{gtrans}.
$U(N)$ gauge theory  is generalised to the NC case along the same lines. Matter fields are  represented by complex  ${\mathcal A}$ modules $\modul=\C^N\otimes \mathcal{A}$ while gauge transformations are automorphisms of $\modul$ which may be realised as Lie algebra valued $\star$-exponentials according to
\be
g(f)= \exp_{\star}\left(i f^j e_j\right), \;\;\; j=1,...,N
\ee
with $f^j\in \mathcal{A}$ and $e_j$ a Hermitian  basis in $\C^N$.

\end{appendix}

\end{document}